%% file: hep-ph.tex
\documentclass{article}

\usepackage{epsfig}
\usepackage{amsmath}
\usepackage{xspace}
\usepackage{cite}

\newcommand{\LLx}{LL$x$\xspace}
\newcommand{\NLLx}{NLL$x$\xspace}

\newcommand{\ie}{i.e.\xspace}
\newcommand{\eg}{e.g.\xspace}
\newcommand{\cnf}{cf.\ }

\newcommand{\NLLB}{NLL$_\mathrm{B}$\xspace}
\newcommand{\as}{\alpha_s}              % coupling constant
\newcommand{\asb}{\bar{\alpha}_s}       % alpha_s bar
\newcommand{\order}[1]{\mathcal{O}\left(#1\right)}

\newcommand{\om}{\omega}

\newcommand{\NC}{N_c}

\newcommand{\cK}{{\cal K}}

\begin{document}

%\markboth{Gavin P. Salam}
%{Fall and rise of the gluon splitting function}

%\catchline{}{}{}{}{}
%
%%%%%%%%%%%%%%%%%%%%%%%%%%%%%%%%%%%%%%%%%%%%%%%%%%%%%%%%%%%%%%%%%%%%

\title{\Large \textsc{Fall and rise of the gluon splitting
    function}\footnote{Talk presented at DIS 2004, \v{S}trbsk\'e Pleso,
    Slovakia, April 2004, and at the Eighth Workshop on
    Non-Perturbative Quantum Chromodynamics, Paris, France, June 2004.}}

\author{Gavin P. Salam\smallskip\\
LPTHE, Universities of Paris VI \& VII and CNRS,\\
75252 Paris 75005, France.
}

%\address{}

\date{}

\maketitle

{\vspace{-5cm}
\begin{flushright}
  LPTHE--P04--04\\
  hep-ph/0407368
\end{flushright}
\vspace{3.0cm}}

%\pub{Received (Day Month Year)}{}

\begin{abstract}
This talk reviews some recent results on the NLL resummed
  small-$x$ gluon splitting function, as determined including
  renormalisation-group improvements. It also discusses the
  observation that the LO, NLO, NNLO, etc.\ hierarchy for the gluon
  splitting function breaks down not when $\as \ln 1/x \sim 1$ but
  rather for $\as \ln^2 1/x \sim 1$.
%
%\keywords{High-energy limit of QCD; BFKL; splitting functions.}
\end{abstract}
 
\input{salam_fall_rise.tex}

\end{document}

%% file: salam_fall_rise.tex
It is well-known that the resummation of leading-logarithms of $x$
(\LLx) \cite{BFKL} and next-to-leading logarithms (\NLLx)
\cite{NLLFL,NLLCC} in the gluon splitting function,
\begin{equation}
  \label{eq:Pggstruct}
  xP_{gg}(x) = \sum_{n=1} A_{n,n-1} \,\asb^{n}\, \ln^{n-1} \frac1x
   + \sum_{n=2} A_{n,n-2}\,\asb^{n}\, \ln^{n-2} \frac1x +
   \dots\,,\quad
   \asb = \frac{\as\NC}{\pi},
\end{equation}
leads to strong small-$x$ growth (positive at \LLx, negative at
\NLLx), quite inconsistent with data \cite{EHW,BF95}. In contrast,
fits to data with purely NLO splitting functions, free of any
small-$x$ enhancements (because $A_{21}=0$), work remarkably well
\cite{CTEQNLO,MRSTNLO} even at $x\sim 10^{-4}$, where one would
expect the effects of resummation to be relevant.

This apparent paradox, and related issues, have been the subject of
intense investigation over the past few years
\cite{ABF2000,ABF2001,ABF2003,ABFcomparison,THORNE,Salam1998,CC,CCS1,CCS00,CCSSkernel,SCHMIDT,FRSV},
and have become all the more relevant with the recent calculation of
the NNLO splitting functions \cite{NNLO}, which explicitly show the
first of the small-$x$ enhanced terms, $A_{31}$ (the \LLx $A_{32}$
term is also zero).

A characteristic of nearly all such approaches is that the resummed
$P_{gg}$ splitting function ends up bearing a strong similarity to the
NLO result, the main additional features being a \emph{dip} in the
splitting function at moderately small values of $x\sim 10^{-3}$,
followed by a slow rise at very small $x$.

The slowness of the rise has been understood in
\cite{CCS1,THORNE,ABF2003} as being, in part, a consequence of running
coupling effects, which convert the branch cut of the fixed-coupling
\LLx splitting function into a set of poles with weak,
$\order{\as^2}$, residues. Furthermore the position of the dominant
pole corresponds to a lower small-$x$ power rise than the position of
the corresponding fixed-coupling branch point, the difference in the
power being of order $b^{2/3}\as^{5/3}$, where
$b=(11-\frac{2n_f}{3})/12$. Just as important, is the consistent
all-orders treatment of renormalisation group logarithms in the BFKL
evolution kernel \cite{Salam1998,CC,CCS1}.

The dip, instead, had until recently received less attention, even
though it appeared in all approaches, and is likely, at today's
energies, to be phenomenologically more relevant than the
asymptotically small-$x$ rise.

\begin{figure}[htbp]
  \centering
  \includegraphics[width=0.48\textwidth]{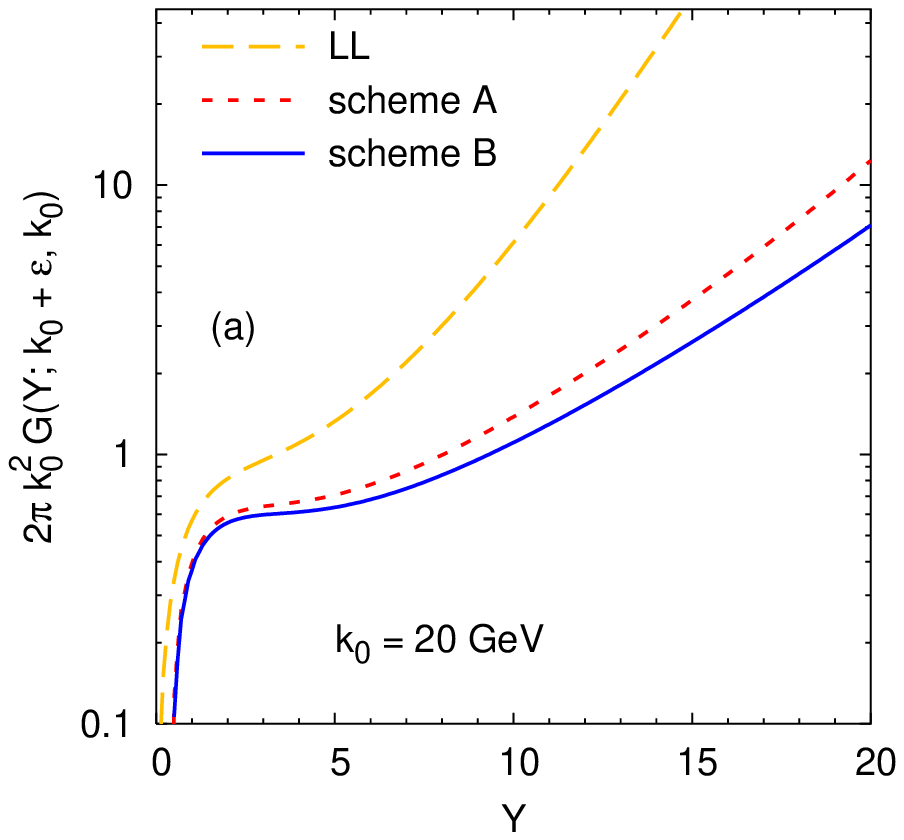}\hfill
  \includegraphics[width=0.48\textwidth]{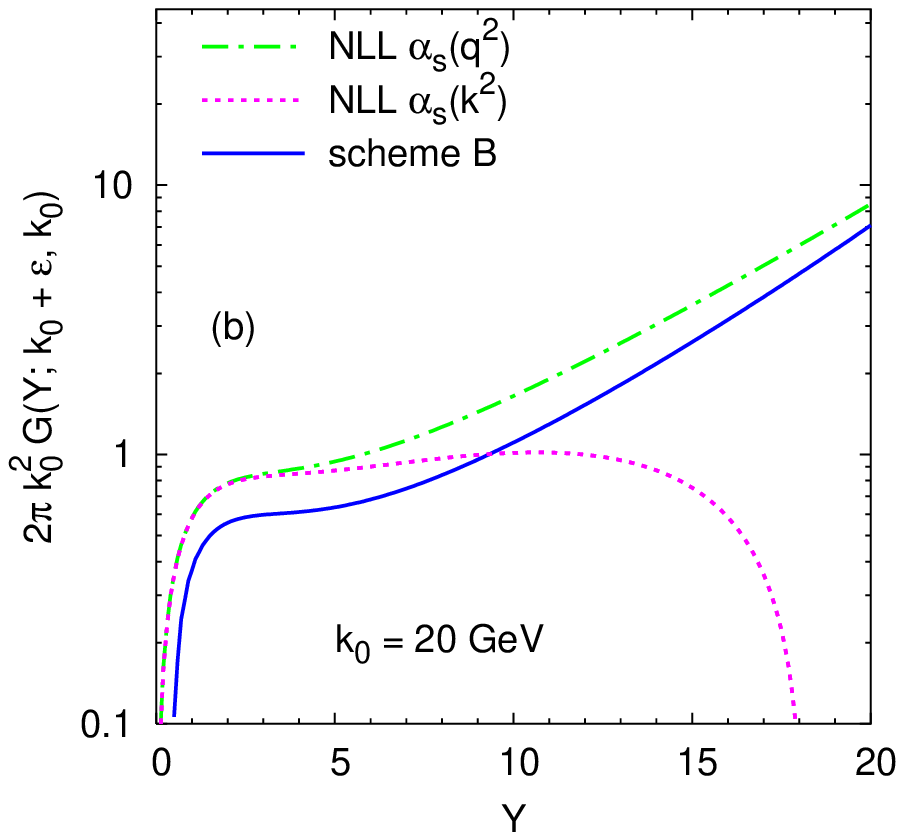}
  \caption{Gluon Green function, as relevant to problems such as
    $\gamma^*\gamma^*$ scattering, with two hard scales, $k$ and
    $k_0$, and $Y\equiv \ln s/k k_0$. Left: compared to \LLx BFKL
    evolution. Schemes A and B represent two forms of RG improvement
    of the \NLLx BFKL kernel.  Right: comparison to pure \NLLx
    evolution with two different schemes for the running of the
    coupling \cite{CCSSkernel}.}
  \label{fig:G}
\end{figure}

Before discussing the dip in detail, it is useful review the methods,
as used in \cite{CCSSkernel}, to extract the resummed small-$x$
splitting function that will be shown here. First one introduces a
kernel $\cK(z,k,k')$ which includes both BFKL (\LLx, \NLLx) and
renormalisation group (RG, \ie LO DGLAP) contributions to small-$x$
evolution. It is designed so as to be symmetric under the exchange of
the two transverse scales, $k_0$ and $k_1$, modulo symmetry-breaking
running-coupling effects. One then numerically determines a gluon
Green function $G(\ln 1/x,k,k_0)$, as the solution of
\begin{equation}
  \label{eq:green-fn-eq}
  G(\ln 1/x,k,k_0) = G_0(\ln 1/x,k,k_0) + \cK \otimes G\,,
\end{equation}
where the inhomogeneous term ($G_0$) represents some arbitrary initial
condition and the convolution $\cK \otimes G$ is over both
longitudinal momenta $x$ and transverse momenta $k$. The Green
function can, among other things, be used as an input for studies of
processes such as $\gamma^*\gamma^*$, in which case one takes $k\sim
k_0 \gg \Lambda_{QCD}$. This is illustrated in fig.\ref{fig:G}, where
the curves labelled `scheme A' and `B' correspond to two variants of
the RG-improved \NLLx kernel. Features to note include the %relatively
slow onset of the BFKL growth of the Green function compared to \LLx
evolution, and the good degree of stability of the RG improved
approach compared to pure \NLLx approaches, the latter being sensitive
both to the details of the treatment of the running coupling as well
as to the precise definition of $Y$ ($\ln s/kk_0$ or $\ln s/k^2$, not
shown).\footnote{Another running coupling scheme for pure \NLLx has
  also been considered in \cite{Andersen-SabioVera}, giving results
  similar to our $\as(q^2)$ curves. It is within an approach that has
  the advantage of providing easy access to information also on the
  dependence of $G(\ln 1/x,\vec k,{\vec k}_0)$
  on the azimuthal angle between $\vec k$ and ${\vec k}_0$} %

\begin{figure}[t]
  \centering
  \includegraphics[width=0.75\textwidth,height=0.55\textwidth]{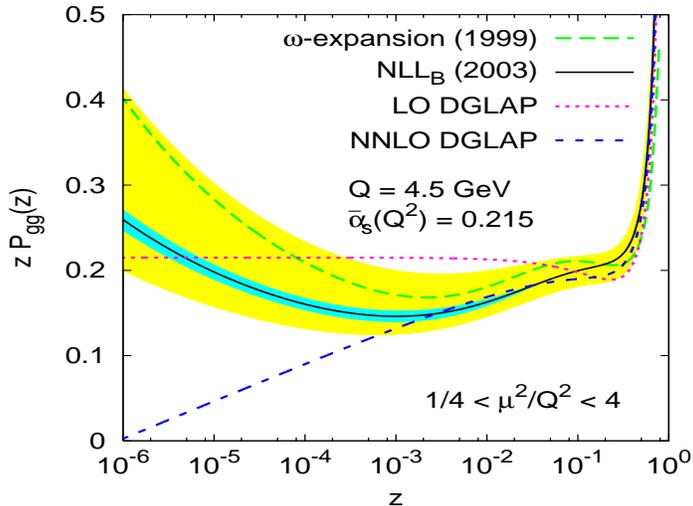}
  \caption{$P_{gg}$ resummed small-$x$ splitting function, as
    determined in \cite{CCSSkernel}, together with a shaded band from
    the variation of the renormalisation scale in the resummation. It
    is compared to an older resummed determination, in the
    $\omega$-expansion \cite{CCS1}, the LO DGLAP splitting function
    and the recently determined NNLO splitting function \cite{NNLO}.}
  \label{fig:split}
\end{figure}

Rather than such same-scale processes, here our main interest is in
cases where $k_0$ is some non-perturbative scale $\sim \Lambda_{QCD}$,
while $k \gg k_0$. One can then introduce a gluon distribution,
\begin{equation}
  x g(x,Q^2) \equiv \int^{Q} d^2k \; G^{(\nu_0=k^2)}(\ln 1/x, k,
  k_0)\,,\qquad
  x = \frac{k^2}{s}\,.
\end{equation}
Perturbatively, $g(x,Q^2)$ is not under control, since it depends on
non-perturbative information such as the initial condition $G^{(0)}$,
as well as the detailed dynamics in the infrared. However, because
collinear factorisation holds even at small-$x$, the splitting
function $P_{gg}(z,Q^2)$ \emph{is} perturbatively defined,
through the following relation,
\begin{equation}
  \frac{d g(x,Q^2)}{d \ln Q^2} = \int \frac{dz}{z}\, {
    P_{gg}(z,Q^2)}\,
   g\left(\frac{x}{z},Q^2\right)\,.
\end{equation}
Thus one can numerically evaluate the gluon distribution for a
particular set of non-perturbative criteria (\eg infrared
cutoff on $\as$) and then carry out the deconvolution so as to obtain
$P_{gg}$. This gives the splitting function of fig.\ref{fig:split}
(\NLLB, corresponding to scheme B of fig.\ref{fig:G}), together with
an inner uncertainty band, associated with an estimate of residual
(higher-twist) dependence on the non-perturbative dynamics. The small
size of the latter serves as a cross-check of factorisation.

Relevant features of the splitting function of fig.\ref{fig:split}
include the dip at moderately small $x$, followed by the slow rise at
much smaller $x$. To within renormalisation scale uncertainties (outer
shaded band) it is compatible with the inverse Mellin transform of the
anomalous dimension determined in \cite{CCS1}, which was based on
similar physical assumptions ($\om$-expansion) but with different
subleading terms, thus confirming the overall consistency between
different approaches. Regarding the potential for future comparisons
with data, an interesting similarity was observed in
\cite{ABFcomparison} with the splitting function of \cite{ABF2000},
which had in part been fitted to the HERA data.

As already mentioned, the main feature of the splitting function in
the phenomenologically relevant region of $x\sim 10^{-3}$ is a dip. It
is important that one understand the origin of the dip, and establish
whether it is a robust small-$x$ prediction or rather some form of
artefact.

A first 
point of interest in this respect concerns the striking
similarity, down to $x\sim10^{-3}$, with the recently calculated
NNLO splitting function \cite{NNLO}.  Since the \LLx $A_{32}$ term is
zero, the NNLO splitting function, at small-$x$, is dominated by a
(\NLLx) term, $A_{31} \asb^3 \ln 1/x$, where $A_{31}$ is negative. It
also contains a N\NLLx term, $A_{30} \asb^3$, but this contributes
only an overall shift to the height of the curve, without modifying
its $x$-dependence. This suggests that for moderately small $x$ it
might be the $A_{31} \asb^3 \ln 1/x$ term that dominates the resummed
splitting function. To see whether this is the case, one has to
examine the relative size of other logarithmically enhanced terms. At
the next order in $\as$, the term with the strongest $x$ dependence is
$A_{43} \asb^4 \ln^3 1/x$. Taking just the $A_{31}$ and $A_{43}$ terms
gives (ignoring constant terms of order $\as^2$),
\begin{equation}
  \label{eq:pgg-approx-first-rtas}
  xP_{gg}(x) \simeq \as + A_{31} \asb^3 \ln 1/x + A_{43} \asb^4 \ln^3 1/x\,,
\end{equation}
which has a minimum at $\ln 1/x =
\sqrt{-\frac{A_{31}}{3A_{43}}\frac1{\asb}}$, since $A_{31}$ and
$A_{43}$ have opposite signs. The appearance here of $1/\sqrt{\asb}$
is quite unexpected, since small-$x$ resummation effects are normally
expected to set in for $\asb\ln 1/x \sim 1$. It is a consequence of
the absence of the $A_{21}\asb^2 \ln 1/x$ and $A_{32}\asb^3 \ln^2 1/x$
\LLx terms.

\begin{figure}[htbp]
  \centering
  \begin{minipage}[c]{0.48\textwidth}
    \includegraphics[width=\textwidth]{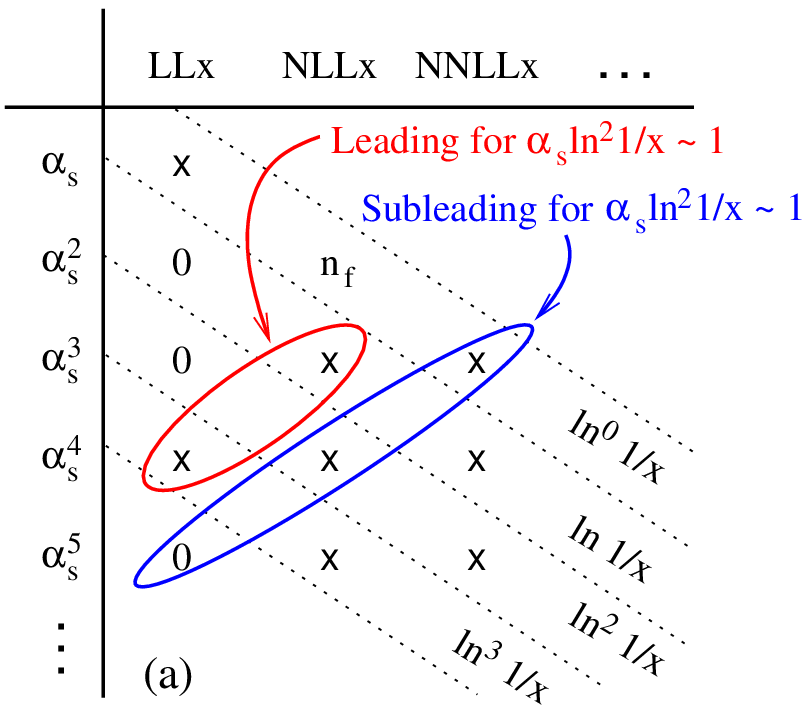}\\
    \mbox{ }
  \end{minipage}
  \hfill
  \begin{minipage}[c]{0.48\textwidth}
    \includegraphics[width=\textwidth]{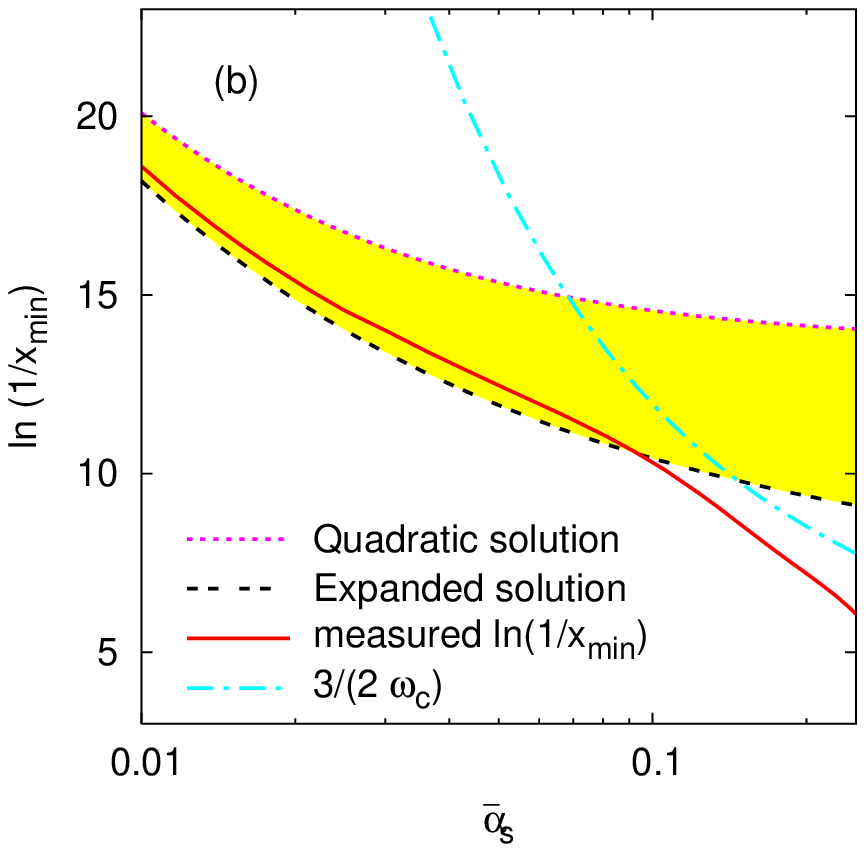}\vspace{-0.3cm}
  \end{minipage}
  \caption{(a) Various classifications of small-$x$ logarithmically
    enhanced terms, $n_f$ indicating that a term is proportional to
    $n_f$. (b) The position of the dip as a function of $\asb$, as
    measured from the numerical calculations of \cite{CCSSkernel}; as
    estimated from the expanded solution, eq.(\ref{eq:xmin}), together
    with an uncertainty band (see \cite{CCSSdip} for details); and
    compared to $3/(2\omega_c)$, an upper bound for $\ln 1/x_{min}$,
    based on resummation arguments.}
  \label{fig:table}
\end{figure}
Having established the existence of two terms leading to a dip
structure for $\as \ln^2 x \sim1$, the next step is to examine the
impact of yet higher-order terms.  Fig.\ref{fig:table}a illustrates
the various terms that are present, highlighting the classification in
the normal \LLx hierarchy, the number of powers of $\as$ and of $\ln
x$.  One sees that eq.(\ref{eq:pgg-approx-first-rtas}), \ie the terms
grouped in the upper (red) ellipse, includes the complete set of terms
of order $\as^{5/2}$ for $\as \ln^2x\sim1$. One can also identify a
finite set of terms (lower, blue ellipse) that contribute at order
$\as^3$ in this region and so forth. This ability to classify all
contributions into finite groups of terms of a common order of
magnitude for $\as\ln^2x\sim1$ means that our identification of a dip
associated with the structure of eq.(\ref{eq:pgg-approx-first-rtas}) is
robust. We can write an expansion for its position \cite{CCSSdip},
\begin{equation}
  \label{eq:xmin}
  \ln \frac{1}{x_{min}} = \sqrt{-\frac{A_{31}}{3A_{43}}\frac1{\asb}}
  - \frac{A_{42}}{3A_{43}} + \order{\sqrt{\asb}}
  \simeq \frac{1.156}{\sqrt{\asb}} + 6.947 + \order{\sqrt{\asb}}\,,
\end{equation}
while its depth $-d$ below the level of the constant $\asb$ and
$\asb^2$ contributions is
\begin{subequations}
  \label{eq:Pggdepth}
\begin{align}
%  D = A_{20} \asb^2 + \frac{2}{9}
%  \sqrt{\frac{-3A_{31}^3}{A_{43}}}\asb^{5/2} - 
%  \left(\frac13 
%  \frac{A_{31}A_{42}}{A_{43}}  + A_{30}\right)\asb^3 + \order{\asb^{7/2}}\,.
  -d &= \frac{2A_{31}}{9} \sqrt{\frac{-3A_{31}}{A_{43}}}\asb^{5/2} -
  \frac13 \frac{A_{31}A_{42}}{A_{43}} \asb^3 + \order{\asb^{7/2}}
  \\&
  \simeq -1.237\asb^{5/2} - 11.15\asb^3 + \order{\asb^{7/2}}\,.
\end{align}
\end{subequations}
Given that the expansion is in powers of $\sqrt{\as}$, the large
coefficients of the higher-order terms (given for $n_f=4$) mean that
one can only trust the expansion for extremely small values of $\as$,
and one may legitimately wonder whether the discussion given here is
of relevance to the dip seen for realistic values of $\as$.

To help answer this question, fig.\ref{fig:table}b shows the position
of the dip as measured from the numerical calculations of
\cite{CCSSkernel} (solid line) compared to eq.(\ref{eq:xmin}) (dashed
line) together with a band indicating the estimated size of
$\order{\sqrt{\as}}$ uncertainties on eq.(\ref{eq:xmin}). There is
remarkably good agreement up to $\asb\simeq 0.1$, despite the fact
that eq.(\ref{eq:xmin}) seems to converge only for much smaller $\as$.

Beyond $\asb\simeq0.1$ one sees a change in behaviour of the position
of the minimum. To understand it, we note that the usual
(fixed-coupling) approximation for the resummed splitting function is
\begin{equation}
  \label{eq:resummed_pgg}
  xP_{gg}(x) \sim \frac{x^{-\omega_c}}{\ln^{3/2} 1/x}\,,
\end{equation}
where $\omega_c$ is the asymptotic power for the growth of
$xP_{gg}(x)$. The point where eq.(\ref{eq:resummed_pgg}) starts to
rise, $\ln 1/x \sim 3/(2\om_c)$, corresponds to that where the
all-orders resummation of $\as \ln 1/x$ terms becomes important.  This
point would be expected to provide an upper bound on the position of
any minimum of $xP_{gg}(x)$, and indeed, plotting it in
fig.\ref{fig:table}b reveals that the kink in the $\asb$-dependence of
$\ln 1/x_{min}$ occurs precisely at the point where the $\ln 1/x \sim
3/(2\om_c)$ bound becomes relevant. What occurs here is that the upper
bound, of order $1/\as$, becomes smaller than the formal position of
the minimum, of order $1/\sqrt{\as}$, as a consequence of the relative
sizes of the (subleading) coefficients of the two series.

%With this information at our disposal, 
Overall, we arrive at the following
overall picture regarding the dip: the initial decrease seems in all
cases, including for realistic values of $\as$, to be due to the
negative $A_{31} \asb^3 \ln1/x$ term. For %sufficiently 
small $\as$,
the decrease continues up to the point, $\ln 1/x \sim 1/\sqrt{\as}$,
where the $A_{43} \asb^4 \ln^3 1/x$ becomes of similar order, \cnf
eq.(\ref{eq:xmin}). But for realistic values of $\alpha_s$, the full
BFKL rise actually sets in earlier than is given by
eq.(\ref{eq:xmin}), roughly for $\ln 1/x \sim 3/(2\om_c)$ and it is
this point that then determines the position of minimum of the dip.
This global picture leads us to believe that the dip is a rigorous
property of the small-$x$ expansion, even if it is difficult to
calculate its properties analytically for phenomenologically relevant
values of $\as$.

\textbf{Acknowledgements.} The work described here has been carried
out in collaboration with M.~Ciafaloni, D.~Colferai and A.~Sta\'sto.